\documentclass[10pt, twocolumn, final]{IEEEtran}%

\usepackage[T1]{fontenc}
\usepackage{amsmath,amssymb,amsfonts,mathrsfs,bm}
\usepackage{mathtools}
\usepackage{amsthm}
\usepackage{nicefrac}
\usepackage{cite}
\usepackage[shortlabels]{enumitem}
\usepackage{graphicx}
\usepackage{epstopdf}
\usepackage{url}
\usepackage{colortbl}
\usepackage{booktabs}
\usepackage{multirow}
\usepackage[table,dvipsnames]{xcolor}
\usepackage[normalem]{ulem}

\usepackage{array}
\newcolumntype{L}[1]{>{\raggedright\let\newline\\\arraybackslash\hspace{0pt}}m{#1}}
\newcolumntype{C}[1]{>{\centering\let\newline\\\arraybackslash\hspace{0pt}}m{#1}}
\newcolumntype{R}[1]{>{\raggedleft\let\newline\\\arraybackslash\hspace{0pt}}m{#1}}

\makeatletter
\let\MYcaption\@makecaption
\makeatother
\usepackage[font=footnotesize]{subcaption}
\makeatletter
\let\@makecaption\MYcaption
\makeatother

\usepackage{xparse}



\usepackage{glossaries}

\makeatletter
\let\oldgls\gls
\let\oldglspl\glspl

\newcommand\fussy@ifnextchar[3]{%
  \let\reserved@d=#1%
  \def\reserved@a{#2}%
  \def\reserved@b{#3}%
  \futurelet\@let@token\fussy@ifnch}
\def\fussy@ifnch{%
  \ifx\@let@token\reserved@d
    \let\reserved@c\reserved@a 
  \else
    \let\reserved@c\reserved@b
  \fi
 \reserved@c}

\renewcommand{\gls}[1]{%
  \oldgls{#1}\fussy@ifnextchar.{\@checkperiod}{\@}}
\renewcommand{\glspl}[1]{%
  \oldglspl{#1}\fussy@ifnextchar.{\@checkperiod}{\@}}

\newcommand{\@checkperiod}[1]{%
  \ifnum\sfcode`\.=\spacefactor\else#1\fi
}
\makeatother

\newacronym{wrt}{w.r.t.}{with respect to}
\newacronym{RHS}{RHS}{right-hand side}
\newacronym{LHS}{LHS}{left-hand side}
\newacronym{iid}{i.i.d.}{independent and identically distributed}

\usepackage{float}

\ifx\notloadhyperref\undefined
	\ifx\loadbibentry\undefined
		\usepackage[hidelinks,hypertexnames=false]{hyperref} 
	\else
		\usepackage{bibentry}
		\makeatletter\let\saved@bibitem\@bibitem\makeatother
		\usepackage[hidelinks,hypertexnames=false]{hyperref}
		\makeatletter\let\@bibitem\saved@bibitem\makeatother
	\fi
\else
	\ifx\loadbibentry\undefined
		\relax
	\else
		\usepackage{bibentry}
	\fi
\fi

\usepackage[capitalize]{cleveref}
\crefname{equation}{}{}
\Crefname{equation}{}{}
\crefname{claim}{claim}{claims}
\crefname{step}{step}{steps}
\crefname{line}{line}{lines}
\crefname{condition}{condition}{conditions}
\crefname{dmath}{}{}
\crefname{dseries}{}{}
\crefname{dgroup}{}{}

\crefname{Problem}{Problem}{Problems}
\crefformat{Problem}{Problem~(#2#1#3)}
\crefrangeformat{Problem}{Problems~(#3#1#4) to~(#5#2#6)}

\crefname{Theorem}{Theorem}{Theorems}
\crefname{Corollary}{Corollary}{Corollaries}
\crefname{Proposition}{Proposition}{Propositions}
\crefname{Lemma}{Lemma}{Lemmas}
\crefname{Definition}{Definition}{Definitions}
\crefname{Example}{Example}{Examples}
\crefname{Assumption}{Assumption}{Assumptions}
\crefname{Remark}{Remark}{Remarks}
\crefname{Rem}{Remark}{Remarks}
\crefname{remarks}{Remarks}{Remarks}
\crefname{Appendix}{Appendix}{Appendices}
\crefname{Exercise}{Exercise}{Exercises}
\crefname{Theorem_A}{Theorem}{Theorems}
\crefname{Corollary_A}{Corollary}{Corollaries}
\crefname{Proposition_A}{Proposition}{Propositions}
\crefname{Lemma_A}{Lemma}{Lemmas}
\crefname{Definition_A}{Definition}{Definitions}

\usepackage{crossreftools}
\ifx\notloadhyperref\undefined
\else
	\relax
\fi

\usepackage{algorithm,algorithmic}

\ifx\loadbreqn\undefined
	\relax
\else
	\usepackage{breqn} 
\fi

\ifx\renewtheorem\undefined
\ifx\useTheoremCounter\undefined
\newtheorem{Theorem}{Theorem}
\newtheorem{Corollary}{Corollary}
\newtheorem{Proposition}{Proposition}
\newtheorem{Lemma}{Lemma}
\else
\newtheorem{Theorem}{Theorem}

\fi

\newtheorem{Definition}{Definition}
\newtheorem{Example}{Example}

\fi

\theoremstyle{remark}

\theoremstyle{plain}

\newcommand{\Real}{\mathbb{R}}

\newcommand{\calA}{\mathcal{A}}

\newcommand{\calC}{\mathcal{C}}

\newcommand{\calG}{\mathcal{G}}

\newcommand{\calN}{\mathcal{N}}

\newcommand{\calP}{\mathcal{P}}

\newcommand{\calS}{\mathcal{S}}

\newcommand{\calW}{\mathcal{W}}

\newcommand{\bA}{\mathbf{A}}

\newcommand{\bF}{\mathbf{F}}

\newcommand{\bL}{\mathbf{L}}

\newcommand{\bM}{\mathbf{M}}

\newcommand{\bP}{\mathbf{P}}

\newcommand{\bS}{\mathbf{S}}

\newcommand{\bU}{\mathbf{U}}

\newcommand{\bw}{\mathbf{w}}

\newcommand{\bx}{\mathbf{x}}

\newcommand{\by}{\mathbf{y}}

\newcommand{\bz}{\mathbf{z}}

\DeclareSymbolFont{bsfletters}{OT1}{cmss}{bx}{n}
\DeclareSymbolFont{ssfletters}{OT1}{cmss}{m}{n}
\DeclareMathSymbol{\bsfGamma}{0}{bsfletters}{'000}
\DeclareMathSymbol{\ssfGamma}{0}{ssfletters}{'000}
\DeclareMathSymbol{\bsfDelta}{0}{bsfletters}{'001}
\DeclareMathSymbol{\ssfDelta}{0}{ssfletters}{'001}
\DeclareMathSymbol{\bsfTheta}{0}{bsfletters}{'002}
\DeclareMathSymbol{\ssfTheta}{0}{ssfletters}{'002}
\DeclareMathSymbol{\bsfLambda}{0}{bsfletters}{'003}
\DeclareMathSymbol{\ssfLambda}{0}{ssfletters}{'003}
\DeclareMathSymbol{\bsfXi}{0}{bsfletters}{'004}
\DeclareMathSymbol{\ssfXi}{0}{ssfletters}{'004}
\DeclareMathSymbol{\bsfPi}{0}{bsfletters}{'005}
\DeclareMathSymbol{\ssfPi}{0}{ssfletters}{'005}
\DeclareMathSymbol{\bsfSigma}{0}{bsfletters}{'006}
\DeclareMathSymbol{\ssfSigma}{0}{ssfletters}{'006}
\DeclareMathSymbol{\bsfUpsilon}{0}{bsfletters}{'007}
\DeclareMathSymbol{\ssfUpsilon}{0}{ssfletters}{'007}
\DeclareMathSymbol{\bsfPhi}{0}{bsfletters}{'010}
\DeclareMathSymbol{\ssfPhi}{0}{ssfletters}{'010}
\DeclareMathSymbol{\bsfPsi}{0}{bsfletters}{'011}
\DeclareMathSymbol{\ssfPsi}{0}{ssfletters}{'011}
\DeclareMathSymbol{\bsfOmega}{0}{bsfletters}{'012}
\DeclareMathSymbol{\ssfOmega}{0}{ssfletters}{'012}

\newcommand{\qednew}{\nobreak \ifvmode \relax \else
      \ifdim\lastskip<1.5em \hskip-\lastskip
      \hskip1.5em plus0em minus0.5em \fi \nobreak
      \vrule height0.75em width0.5em depth0.25em\fi}

\newcommand{\ud}{\mathrm{d}}

\newcommand{\norm}[1]{{\left\lVert{#1}\right\rVert}}

\DeclareDocumentCommand\set{s m t| m}{%
  \IfBooleanTF#1%
	{\left\{\, #2\mathrel{} \IfBooleanTF{#3}{\middle|}{:}\mathrel{}  #4\, \right\}}%
  {\{\, #2 \IfBooleanTF{#3}{\mid}{\mathrel{} : \mathrel{}} #4\, \}}%
}

\DeclareDocumentCommand \ifcond {m m} {%
	{#1} %
	\IfValueT{#2}{\, \middle|\, {#2}}%
}

\DeclareDocumentCommand \P {e{_} g >{\SplitArgument{ 1 }{ @| }}d() g } {%
	\mathbb{P}%
	\IfValueTF{#1}{_{#1}}
		{\IfValueT{#2}{_{#2}}}%
	\IfValueT{#3}{\left(\ifcond#3}%
	\IfValueT{#4}{\, \middle|\, {#4}}%
	\IfValueT{#3}{\right)}%
}

\DeclareDocumentCommand \E {e{_} g >{\SplitArgument{ 1 }{ @| }}o g } {%
	\mathbb{E}%
	\IfValueTF{#1}{_{#1}}
		{\IfValueT{#2}{_{#2}}}%
	\IfValueT{#3}{\left[\ifcond#3}%
	\IfValueT{#4}{\, \middle|\, {#4}}%
	\IfValueT{#3}{\right]}%
}

\definecolor{gray90}{gray}{0.9}

\ifx\nohighlights\undefined

	\newcommand{\msout}[1]{\text{\color{green} \sout{\ensuremath{#1}}}}
	\newcommand{\del}[1]{{\color{green}\ifmmode \msout{#1}\else\sout{#1}\fi}}
\else

	\newcommand{\msout}[1]{#1}
	\newcommand{\del}[1]{#1}
\fi

\newcommand{\hhide}[1]{}

\renewcommand{\figurename}{Fig.}
\newcommand{\figref}[1]{\figurename~\ref{#1}}
\graphicspath{{./Figures/}} 
\newcommand{\includeCroppedPdf}[2][]{%
    \IfFileExists{./Figures/#2-crop.pdf}{}{%
        \immediate\write18{pdfcrop ./Figures/#2 ./Figures/#2-crop.pdf}}%
    \includegraphics[#1]{./Figures/#2-crop.pdf}}

\ifx\diagnoselabel\undefined
	\relax
\else
	\makeatletter
	 \def\@testdef #1#2#3{%
		 \def\reserved@a{#3}\expandafter \ifx \csname #1@#2\endcsname
		\reserved@a  \else
	 \typeout{^^Jlabel #2 changed:^^J%
	 \meaning\reserved@a^^J%
	 \expandafter\meaning\csname #1@#2\endcsname^^J}%
	 \@tempswatrue \fi}
	\makeatother
\fi

\crefname{question}{question}{questions}

\usepackage{tikz}
\usepackage{pgfplots}
\pgfplotsset{compat=1.5}

\providecommand{\U}[1]{\protect\rule{.1in}{.1in}}

\theoremstyle{definition}

\begin{document}
\date{}
\title{On distributional graph signals}
\author{Feng~Ji, Xingchao~Jian and Wee~Peng~Tay,~\IEEEmembership{Senior Member,~IEEE} %
\thanks{The authors are with the School of Electrical and Electronic Engineering, Nanyang Technological University, 639798, Singapore (e-mail: jifeng@ntu.edu.sg, xingchao001@e.ntu.edu.sg, wptay@ntu.edu.sg).}%
}
\maketitle

\begin{abstract}
Graph signal processing (GSP) studies graph-structured data, where the central concept is the vector space of graph signals. To study a vector space, we have many useful tools up our sleeves. However, uncertainty is omnipresent in practice, and using a vector to model a real signal can be erroneous in some situations. In this paper, we want to use the Wasserstein space as a replacement for the vector space of graph signals, to account for signal stochasticity. The Wasserstein is strictly more general in which the classical graph signal space embeds isometrically. An element in the Wasserstein space is called a distributional graph signal. On the other hand, signal processing for a probability space of graphs has been proposed in the literature. In this work, we propose a unified framework that also encompasses existing theories regarding graph uncertainty. We develop signal processing tools to study the new notion of distributional graph signals. We also demonstrate how the theory can be applied by using real datasets.
\end{abstract}

\section{Introduction}
Graph signal processing (GSP) is a rapidly growing field that studies signals defined on graphs \cite{Shu12, Shu13, San13, San14, Gad14, Don16, Def16, Kip16, Egi17, Gra18, Ort18, Girault2018, Ji19}. Many real-world phenomena can be naturally represented as graphs, such as social networks, transportation systems, and sensor networks. In GSP, the central concept is the vector space of graph signals, and a graph signal assigns a number to each node of a given graph. Being a vector space, we can use linear transformations, such as the graph Fourier transform and graph filters, to analyze graph signals and study relations among them.  

However, in many practical applications, uncertainty is ubiquitous, and using a vector to model a real signal can be erroneous. The vector space of graph signals assumes that the signal is known exactly, but this is often not the case in real-world scenarios. For example, in a social network, the exact values of the attributes such as user ratings of each user may not be known \cite{Hua20}, or in a sensor network, the sensor readings may be uncertain due to measurement errors or sensor variability \cite{Pra09}. Moreover, it is studied in \cite{Ji23} that in graph neural networks (GNNs), interpreting class labels of nodes as a graph signal can easily ignore label prediction uncertainty and the resulting step graph signal can be highly non-smooth.  

To address this issue, we propose to use the Wasserstein space \cite{Vil09} as a replacement for the vector space of graph signals. An element in the Wasserstein space is a probability distribution on the classical graph signal space. We call such a distribution a distributional graph signal. Therefore, uncertainty is encoded in a distributional graph signal. This provides a more flexible and realistic approach to modeling signals on graphs, which can account for uncertainty and stochasticity. Moreover, the Wasserstein space is strictly more general than the classical vector space of graph signals in which it embeds isometrically. This means that distributional graph signals can accommodate all signals that can be represented as a vector in the classical sense, and more, which can be represented by a probability distribution. By considering distributional graph signals, we can develop a more comprehensive and accurate framework for studying graph-structured data that accounts for uncertainty. In the context of GNN, the notion of distributional graph signals is introduced in \cite{Ji23} and further studied in \cite{Ji23b}. The distributional version of total variation and signal non-uniformity are introduced to enhance the performance of GNNs. In this paper, we want to propose a signal processing framework for distributional graph signals. 

On the other hand, \cite{Ji22} proposes a signal processing framework for a probability space of graph shift operators (see also \cite{Jia23} for an overview), to address the issue that there may not be a single fixed graph topology in many applications. Therefore, in addition to introducing the use of the Wasserstein space for modeling graph signals, we also propose a unified framework that encompasses existing theories regarding graph uncertainty. For this, we introduce the notion of signal adaptive graph structures that associates a distribution of graphs with any graph signal, so that we can construct transformations between distributional graph signals. 

In summary, we replace classical graph signals with distributional graph signals and substitute graph topology with signal adaptive graph structures. As a result, we have a flexible framework to deal with uncertainties in both signals and graphs. In terms of methodology, we have to part from linear algebra and make more use of analysis and probability theory. Therefore, our approach has the flavor of classical Fourier theory \cite{Rud87} rather than that of algebraic signal processing \cite{Pus08}.

Our main contributions are as follows:
\begin{itemize}
    \item We introduce distributional graph signals and signal adaptive graph structures. We develop a signal processing framework by focusing on filter construction.
    \item We relate the framework and the notion of conditional expectation. This allows us to justify some key concepts introduced in \cite{Ji22}.
    \item We explain how classical GSP notions, such as the graph Fourier transform \cite{Shu12, Shu13, Ort18}, convolution \cite{Shu13, Ort18}, and sampling \cite{Aga13, Che15, Tsi15, Mar16, Anis2016, JGT20, Tan20, Lua21}, can be interpreted using the new framework. We use examples to demonstrate that the classical notions are special cases of their counterparts introduced in the paper.
    \item We demonstrate the practical utility of our proposed framework by using real datasets. We show how the proposed approach can be used to analyze and process graph signals with uncertain or stochastic properties. We provide experimental results that demonstrate the effectiveness of the proposed framework.
\end{itemize}

The rest of the paper is organized as follows: In \cref{sec:wsa}, we introduce the Wasserstein space and define the concept of distributional graph signals. In \cref{sec:bay}, we first introduce the notion of signal adaptive graph structures to account for uncertainty in graph topology. Then we explain how they can be used to define transformation between distributional graph signals. The framework is related to the theory of conditional expectation in \cref{sec:coe}. In \cref{sec:gtr}, we review classical GSP theories and describe why they are special cases of the proposed framework. We present numerical results \cref{sec:er} and conclude in \cref{sec:con}. Proofs of all results are deferred to Appendix~\ref{sec:pot}.  

\emph{Notations:} We use $\circ$ to denote function composition. Let $\Real$ denote the set of real numbers and  $M_n(\Real)$ be the space of $n\times n$ real matrices. $\E$ is the expectation operator. Letters $\mu, \nu, \gamma$ are used for probability distributions, while $\delta$ is for delta distributions. We use $\mathcal{A}$ for signal adaptive graph structures (SAGS) introduced in the paper. $G$ is used exclusively for graphs and $f$ is used exclusively for filters. Letters in fraktur font such as $\mathfrak{c}, \mathfrak{p}$ are used to denote a pair of SAGS and a filter. Linear operators and vectors are boldfaced. 

\section{Wasserstein space and distributional graph signals} \label{sec:wsa}

Let $V$ be a set of nodes in a network of size $|V|=n$. A classical signal on $V$ assigns a number to each node of $V$. If an ordering of nodes in $V$ is fixed as $V=\{v_1,\ldots,v_n\}$, then a classical signal can be identified with $\bx\in \mathbb{R}^n$ with the $i$-th component the number assigned to $v_i$. In this paper, we are interested in a probabilistic framework. A natural way to interpret a classical signal $\bx$ is to view it as $\delta_{\bx}$, the delta distribution on $\bx$. This prompts the following generalization of classical signals in terms of the Wasserstein space \cite{Vil09}.

\begin{Definition}
Let $X$ be a metric space. Define the \emph{Wasserstein space} $\calP(X)$ to be the space of (Borel) probability distributions on $X$ with finite mean and variance. If $X =\mathbb{R}^n$, the space of classical graph signals on $V$, then $\calP(X)$ is called the space of \emph{distributional graph signals} on $V$.
\end{Definition}

The main insight is that a distributional signal encodes uncertainties due to reasons such as limitations in measurement precision, forecasting errors, and data labeling mistakes. Hence, using distributional signals can be more realistic than classical signals. The trade-off is that simple and effective tools such as linear algebra are no longer available. In this paper, we shall develop signal processing tools using mainly probability theory and analysis. In view of this, we give $\calP(X)$ a metric \cite{Vil09}.

\begin{Definition}
    Let $X$ be a metric space (with metric $d_X$) and $\calP(X)$ be the associated Wasserstein space. Given $\mu_1,\mu_2$ in $\calP(X)$, the \emph{Wasserstein metric}\footnote{Strictly speaking, the metric considered is the $2$-Wasserstein metric and $2$ accounts for the power in the integral. As this is the only version used in the paper, we omit the quantifier $2$.} $W(\mu_1,\mu_2)$ between $\mu_1,\mu_2$ is defined by
\begin{align*}
    W(\mu_1,\mu_2)^2 = \inf_{\gamma\in \Gamma(\mu_1,\mu_2)}\int d (x,y)^2 \ud\gamma(x,y),
\end{align*}
where $\Gamma(\mu_1,\mu_2)$ is the set of \emph{couplings} of $\mu_1,\mu_2$, i.e., the collection of probability measures on $X \times X$ whose marginals are $\mu_1$ and $\mu_2$, respectively.
\end{Definition}

Intuitively, the Wasserstein metric is the minimum amount of ``work'' required to transform one probability distribution into the other, where the ``work'' is the sum of the product of the amount of probability mass to be moved and the distance that it must be moved. It is well-known that $W(\cdot,\cdot)$ makes $\calP(X)$ a metric space \cite{Vil09}. For distributional graph signals, $\calP(\mathbb{R}^n)$ is complete and separable with the Wasserstein metric.   
It is usually challenging to compute the Wasserstein metric for arbitrary $\mu_1,\mu_2$. However, in special cases, we have closed-form formulas as in the following examples.

\begin{Example} \label{eg:imd}
\begin{enumerate}[(a)]
    \item \label{it:imd} If $\mu_2 = \delta_y$, the delta distribution on $y\in X$, then we have the explicit formula 
    \begin{align*}
           W(\mu_1,\delta_y)^2 = \int d_X(x,y)^2 \ud\mu_1(x).
    \end{align*}
    As a special case, if $\mu_1 = \delta_x$ is also a delta distribution, then $W(\delta_x,\delta_y) = d_X(x,y)$. This implies that the space of classical graph signals $\mathbb{R}^n$ embeds isometrically in the space of distributional graph signals $\calP(\mathbb{R}^n)$. 
    \item Let $\mu_1= \calN(x_1,\Sigma_1)$ and $\mu_2= \calN(x_2,\Sigma_2)$ be two non-degenerate normal distributions on $\mathbb{R}^n$ with mean $x_1,x_2$ and covariance matrices $\Sigma_1,\Sigma_2$ respectively. Then the Wasserstein metric is given by
    \begin{align*}
        W(\mu_1,\mu_2)^2 & = \norm{x_1-x_2}^2\\ & +\text{trace}\big(\Sigma_1+\Sigma_2-2(\Sigma_2^{1/2}\Sigma_1\Sigma_2^{1/2})^{1/2}\big).
    \end{align*}
    Therefore, fitting data in terms of the Wasserstein metric requires one to consider fitting covariance in addition to fitting the mean.
\end{enumerate}
\end{Example}

We have introduced the fundamental object $\calP(\mathbb{R}^n)$ to be studied in the paper. However, we have not yet described contributions from graphs. We explain how graphs enter into the overall picture in the next section.

\section{The Bayesian perspective of distributional graph operators} \label{sec:bay}

In this section, we want to introduce a signal processing framework for distributional graph signals that generalizes classical GSP. There are two aspects of the framework: (1) to encode graph structural information, and (2) to describe (distributional) signal transformations. Each topic occupies one of the following subsections. For concreteness, we do not present the theory in full generality. A more general framework is briefly outlined in Appendix~\ref{sec:act}. 

\subsection{Signal adaptive graph structures}
Let $\calG_n$ be the set of undirected graphs without multiple edges on (ordered) $n$ vertices $V=\{v_1,\ldots,v_n\}$. The graphs can be weighted. Therefore, there is an embedding of $\calG_n$ in $M_n(\mathbb{R})$, the space matrices of size $n$. More specifically, the embedding associates a $G\in \calG_n$ with its weighted adjacency matrix $\bA_G$, where the $(i,j)$-th entry of $\bA_G$ is the weight between $v_i$ and $v_j$. As $M_n(\mathbb{R})$ is measurable with Lebesgure $\sigma$-algebra, it induces a $\sigma$-algebra on $\calG_n$. Moreover, $\calG_n$ is equipped with the subspace topology.

The key insight is that we allow the graph structure to depend on the signal, moreover, it can be random. We formally introduce the following notion.

\begin{Definition}
A \emph{signal adaptive graph structures} (SAGS) assigns to each $\bx \in \mathbb{R}^n$ a probability distribution $\nu_{\bx}$ on $\calG_n$. Denote it by $\calA = (\nu_{\bx})_{\bx\in \mathbb{R}^n}$.   
\end{Definition}

To associate the notion with the Bayesian theory, consider the product space $\mathbb{R}^n \times \calG_n$. It is a measurable space with the product $\sigma$-algebra. The probability distribution $\nu_{\bx}$ can be interpreted as the (conditional) distribution on $\calG_n$ given $\bx\in \mathbb{R}^n$. Therefore, for any distributional graph signal $\mu\in \calP(\mathbb{R}^n)$, we have the an associated distribution $\calA^*(\mu)$ on $\mathbb{R}^n\times \calG_n$ defined by
\begin{align*}
    \calA^*(\mu)(g) = \int \int g(\bx,G) \ud\nu_{\bx}(G)\ud \mu(\bx),
\end{align*}
for any compactly supported continuous function $g$ on $\mathbb{R}^n\times \calG_n$. The distribution $\calA^*(\mu)$ is uniquely determined by the integral formula by the Riesz–Markov–Kakutani representation theorem \cite{Rud87}. The expression reminds us of the law of total probability if $\nu_{\bx}$ is interpreted as the conditional distribution. We now give some examples.

\begin{Example} \label{eg:inn}
\begin{enumerate}[(a)] 
\item \label{it:inn} If $\nu_{\bx} = \nu$, i.e., independent of $\bx$, then we have the setup of \cite{Ji22}. We call it a \emph{constant SAGS}. Moreover, if $\nu=\delta_G$ for a single $G\in \calG_n$, we recover the classical GSP. A further generalization is given next.
\item \label{it:agc} A SAGS $\calA=(\nu_{\bx})_{\bx\in\mathbb{R}^n}$ is \emph{locally constant} for almost every $\bx \in \mathbb{R}^n$, there is an open neighborhood $U_{\bx}$ of $\bx$ such that for every $\by\in U_{\bx}$, we have $\nu_{\by}=\nu_{\bx}$. Intuitively, for such an $\calA$, the signal space $\mathbb{R}^n$ can be (almost) partitioned into open subsets on each of which $\calA$ is a constant. 
\end{enumerate}
\end{Example}

Analogous to these examples, a SAGS $\calA$ encodes the graph structural information. It tells us for a given signal $\bx$, the most suitable graph structures on $V$, according to $\nu_{\bx}$, to process $\bx$. In the next subsection, we describe how distributional signal transformation is performed in this framework.

\subsection{Distributional signal transformations}
Recall that in classical GSP, given a graph $G$, one constructs linear transformations or filters by using the structure of $G$. For example, one may first fix a graph shift operator GSO $\bS$ such as the adjacency matrix or the Laplacian of $G$. Then one applies an algebraic construction such as taking polynomials in $\bS$ to construct desired filters $\bF$. The entire process $G \mapsto \bF$ can be summarized as a map from $\calG_n$ to $M_n(\mathbb{R})$ if we omit the intermediate steps. 

Based on this prototype, we call any measurable function $f: \calG_n \to M_n(\mathbb{R})$ a \emph{pre-filter} or a \emph{pre-transformation}. It induces a measurable function $\widetilde{f}: \mathbb{R}^n\times \calG_n \to \mathbb{R}^n$ by $\widetilde{f}(\bx,G) = f(G)(\bx)$, using the fact that $f(G)\in M_n(\mathbb{R})$ and  $f(G)(\bx)$ is the ordinary matrix operation.  

Given any probability distribution $\mu$ on $\mathbb{R}^n\times \calG_n$, the map $\widetilde{f}$ induces the \emph{pushforward} distribution $f_*(\mu)$ on $\mathbb{R}^n$. More specifically, for any measurable subset $U$ of $\mathbb{R}^n$, we have
\begin{align} \label{eq:fmu}
    f_*(\mu)(U) = \mu(\widetilde{f}^{-1}(U)). 
\end{align}
We do not yet call $f$ a filter or a transformation because we want to impose more constraints on $f$ regarding the distributional graph signals $\calP(\mathbb{R}^n)$.

\begin{Definition} \label{defn:gas}
    Given an SAGS $\calA$, a measurable $f: \calG_n \to M_n(\mathbb{R})$ is a \emph{filter} or a \emph{transformation} with respect to (w.r.t.) $\calA$ if for any distributional graph signal $\mu\in \calP(\mathbb{R}^n)$, the distribution $f_*\circ\calA^*(\mu)$ is also a distributional graph signal, i.e., $f_*\circ\calA^*(\mu)\in \calP(\mathbb{R}^n)$. For convenience, we use $\mathfrak{c}$ to denote the pair $(\calA,f)$ and write $\mathfrak{c}_*(\mu)$ for $f_*\circ\calA^*(\mu)$, if no confusions arise.   
\end{Definition}

The map $\mathfrak{c}_*(\mu): \calP(\mathbb{R}^n) \to \calP(\mathbb{R}^n)$ satisfies the following explicit integral formula:
\begin{align*}
\mathfrak{c}_*(\mu)(g) = 
    \int \int g\big(f(G)(\bx)\big) \ud\nu_{\bx}(G)\ud \mu(\bx),
\end{align*}
for any compactly supported continuous function on $\mathbb{R}^n$.

As we have mentioned, the space distributional graph signals $\calP(\mathbb{R}^n)$ is not linear and we want to focus on the analytic perspective of filters. Recall that one of the most desired analytic properties of a linear map (in functional analysis) is continuity, or equivalently boundedness \cite{Lax02}. In our framework, we also want to study when the map $\mathfrak{c}_*(\mu): \calP(\mathbb{R}^n) \to \calP(\mathbb{R}^n)$ induced by a filter $f$ is continuous. 

For this, we notice that $\calA$ and $f$ give rise to a probability distribution $f_{\calA}(\bx)$ on $M_n(\mathbb{R})$ given $\bx \in \mathbb{R}^n$. Recall $\calA = (\nu_{\bx})_{\bx\in \mathbb{R}^n}$, and the distribution $f_{\calA}(\bx)$ is given by 
\begin{align*}
    f_{\calA}(\bx)(U)=f_*(\nu_{\bx})(U)=\nu_{\bx}(f^{-1}(U)),
\end{align*}
for any measurable subset $U$ of $M_n(\mathbb{R})$. Intuitively, the SAGS $\calA$ associates a family of probable graphs (according to $\nu_{\bx}$) to each $\bx \in \mathbb{R}^n$ and the filter $f$ turns them into a family of probable linear maps that in terms of $f_{\calA}(\bx)$. We endow $M_n(\mathbb{R})$ with the operator norm.

\begin{Theorem} \label{thm:lkb}
Let $K$ be a compact subset of $\mathbb{R}^n$. If $f_{\calA}$ (when restricted to $K$) is a continuous function from $K$ to $\calP(M_n(\mathbb{R}))$, then restricted to $\calP(K)$, $\mathfrak{c}_*: \calP(K) \to \calP(\mathbb{R}^n)$ is uniformly continuous.
\end{Theorem}

In many practical situations, it is reasonable to assume that signals belong to a compact and hence bounded subset of $\mathbb{R}^n$. In such a case, the condition of the theorem is not restrictive. We also have a version of the continuity result without the compactness assumption.

\begin{Theorem} \label{thm:imi}
If $f_{\mathcal{A}}$ is Lipschitz continuous, then $\mathfrak{c}_*(\mu)$ is continuous at any $\mu \in \mathcal{P}(\mathbb{R}^n)$ with finite $6$-th moments. 
\end{Theorem}

We remark that the condition on finite $6$-th moment can be further improved. However, it is sufficient for us as it already includes essential cases such as compactly supported distributions and (mixed) Gaussian distributions. 

We have the following consequence of the result. It is known (e.g., \cite{Vil09}) that finite point distributions are dense in $\mathcal{P}(\mathbb{R}^n)$, i.e, for any distributional graph signals $\mu\in \mathcal{P}(\mathbb{R}^n)$, there is a sequence $(\mu_i)_{i\geq 1}$ of distributional graph signals each supported on finitely many points such that $\mu_i \to \mu, i\to \infty$. If $\mu$ has bounded $6$-th moment and $\mathcal{X}$ satisfies the conditions of \cref{thm:imi}, then by continuity, we have $\mathfrak{c}_*(\mu_i) \to \mathfrak{c}_*(\mu), i\to \infty$. This means that knowledge of the filter at delta distributions tells us a lot about the filter at more general distributions. 

\section{Conditional expectations} \label{sec:coe}
In this section, we propose construction based on a $\mathfrak{c}$ that is related to conditional expectations \cite{Kol56}. The approximation result \cref{thm:ftf} justifies many constructions in \cite{Ji22}. 

As we have seen in the previous section, given a pair of SAGS and a filter $\mathfrak{c}=(\calA,f)$, we have $\mathfrak{c}_*=f_*\circ \calA^*: \calP(\mathbb{R}^n) \to \calP(\mathbb{R}^n)$. On the other hand, for any measurable function $g:\mathbb{R}^n \to \mathbb{R}^n$ and $\mu \in \calP(\mathbb{R}^n)$, pushforward (cf.\ (\ref{eq:fmu})) induces a probability distribution $g_*(\mu)$ on $\mathbb{R}^n$. We call $g$ \emph{bounded} if $g_*(\mu) \in \calP(\mathbb{R}^n)$ for any $\mu \in \calP(\mathbb{R}^n)$. Denote the set of bounded measurable functions by $B(\mathbb{R}^n)$. For example, a linear transformation is bounded and hence belongs to $B(\mathbb{R}^n)$. 

In classical GSP when $\calA$ is a constant delta distribution, $\mathfrak{c}_*$ is induced by the pushforward of a linear transformation. It is easier to study such a map coming directly from a function on the more familiar space $\mathbb{R}^n$. However, for a general $\mathfrak{c}= (\calA,f)$, it is not always true that $\mathfrak{c}_* = g_*$ for some $g\in B(\mathbb{R}^n)$. Nevertheless, it is possible to find good approximations of $\mathfrak{c}_*$. For this, we introduce a function $e_{\mathfrak{c}}$ as follows.

 To construct $e_{\mathfrak{c}}$, assume that $\calA = (\nu_{\bx})_{\bx\in \mathbb{R}^n}$. For $\bx \in \mathbb{R}^n$, we define 
\begin{align} \label{eq:ebi}
e_{\mathfrak{c}}(\bx) = \int f(G)(\bx) \ud\nu_{\bx}(G) = \int\by\ud \mathfrak{c}_*(\delta_{\bx})(\by).
\end{align}
It is related to conditional expectation as follows. Let $p: \mathbb{R}^n\times \calG_n \to \mathbb{R}^n$ be the projection to the first component. Recall that for any $\mu \in \mathcal{P}(\mathbb{R}^n)$, we have constructed the distribution $\calA^*(\mu)$ on $\mathbb{R}^n\times \calG_n$. In this respect, both $p: \mathbb{R}^n\times \calG_n \to \mathbb{R}^n$ and $f: \mathbb{R}^n\times \calG_n \to \mathbb{R}^n$ can be viewed as random variables on the sample space $\mathbb{R}^n\times \calG_n$. It is well known that there is a condition expectation $e_{f}: \mathbb{R}^n \to \mathbb{R}^n$ such that $e_{f}(\bx) = e_{\mathfrak{c}}(\bx)$ up to a set with $\mu$ measure $0$. Due to this fact, the promised approximation property of $e_{\mathfrak{c}}$ reads as follows.

\begin{Theorem} \label{thm:ftf}
For $\mathfrak{c}=(\calA,f)$, the function $e_{\mathfrak{c}}$ is measurable and belongs to $B(\mathbb{R}^n)$. Moreover, for any $g \in B(\mathbb{R}^n)$ and subset $S \subset \mathbb{R}^n$, the following holds:
\begin{align*}
   \sup_{\text{supp}(\mu) \subset S} W\big(e_{\mathfrak{c},*}(\mu), \mathfrak{c}_*(\mu)\big) \leq \sup_{\text{supp}(\nu) \subset S} W\big(g_*(\nu), \mathfrak{c}_*(\nu)\big),
\end{align*}
where $W$ is the Wasserstein metric and the supreme is taken over $\mu$ (resp.\ $\nu$) in $\calP(\mathbb{R}^n)$ supported in $S$. 
\end{Theorem}

We give some examples. 

\begin{Example} \label{eg:ici}
If $\calA$ is a constant SAGS with the common probability measure $\nu$ (cf.\ \cref{eg:inn}\ref{it:inn}), then $e_{\mathfrak{c}}$ is the linear transformation given by the operator
\begin{align*}
    e_{\mathfrak{c}}(\cdot) = \int f(G)(\cdot) \ud\nu(G).
\end{align*}
Similarly, if $\calA$ is a locally constant SAGS (cf.\ \cref{eg:inn}\ref{it:agc}), then outside a subset of measure $0$, the function $e_{\mathfrak{c}}$ is piecewise linear, i.e., for each $\bx$ there is an open neighborhood of $\bx$ on which $e_{\mathfrak{c}}$ is linear.  
\end{Example}

The construction of $e_{\mathfrak{c}}$ enjoys other analytic properties. 
\begin{Lemma} \label{lem:cas}
    Consider a sequence $\mathfrak{c}_i = (\calA_i,f_i), i\geq 1$. If there is a $\mathfrak{c}=(\calA,f)$ such that $f_{i,\calA_i}(\bx) \to f_{\calA}(\bx)$ as $i\to \infty$, then $e_{\mathfrak{c}_i}(\bx) \to e_{\mathfrak{c}}(\bx)$.
\end{Lemma}
Intuitively, the lemma says that the construction $e_{\mathfrak{c}}$ is ``continuous'' in $\mathfrak{c}$. 

For the rest of this section, we discuss some algebraic properties of $e_{\mathfrak{c}}$. Unlike classical GSP, Wasserstein spaces are not linear. However, we can still define binary operations such as addition, analogous to the sum of random variables. 

Let $f_1,f_2$ be filters w.r.t.\ SAGSs $\calA_1=(\nu_{1,\bx})_{\bx\in \mathbb{R}^n}$ and $\calA_2 = (\nu_{2,\bx})_{\bx\in \mathbb{R}^n}$ respectively and denote $(\calA_i,f_i)$ by $\mathfrak{c}_i, i=1,2$. We define the addition $\mathfrak{c}_{1,*}\boxplus\mathfrak{c}_{2,*}$ by the property
\begin{align*}
&\mathfrak{c}_{1,*}\boxplus\mathfrak{c}_{2,*}(\mu)(g) = \\ &\int \int g\big(f_1(G_1)(\bx)+f_2(G_2)(\bx)\big) \ud\nu_{1,\bx}\times \nu_{2,\bx}(G_1,G_2)\ud \mu(\bx),
\end{align*}
for any continuous function $g$ with compact support on $\mathbb{R}^n$. The addition $\mathfrak{c}_{1,*}\boxplus\mathfrak{c}_{2,*}$ allows us to combine filters w.r.t.\ different SAGSs. From the expression, we see its similarity to the sum of random variables. Analogous to (\ref{eq:ebi}), its associated ``conditional expectation'' $e_{\mathfrak{c}_1\boxplus\mathfrak{c}_2}$ is given by the integral
\begin{align*}
e_{\mathfrak{c}_1\boxplus\mathfrak{c}_2}(\bx) = \int \by \ud\mathfrak{c}_{1,*}\boxplus\mathfrak{c}_{2,*}(\delta_{\bx})(\by).
\end{align*}

Scalar multiplication is simpler: given $r\in \mathbb{R}$ and $\mathfrak{c}=(\calA,f)$, then $r\mathfrak{c}$ denote the pair $(\calA,rf)$. The construction of $e_{\mathfrak{c}}$ from $\mathfrak{c}$ respects addition and scalar multiplication.

\begin{Lemma} \label{lem:tam}
The addition $\mathfrak{c}_{1,*}\boxplus\mathfrak{c}_{2,*}$ is well defined. Moreover, for any $r\in \mathbb{R}$, we have $e_{r\mathfrak{c}_1\boxplus\mathfrak{c}_2} = re_{\mathfrak{c}_1} + e_{\mathfrak{c}_2}.$
\end{Lemma}

In the next section, we revisit some key concepts of graph signal processing theories and interpret them with the new framework. 

\section{GSP theories revisited} \label{sec:gtr}

In this section, we describe how we may understand some of the most important GSP concepts in the proposed work, in view of how they are perceived before. We mainly base on \cite{Shu12} and \cite{Ji22}, which are briefly reviewed in \cref{eg:inn} and \cref{eg:ici}.

Given a graph $G \in \calG_n$, recall that the Fourier transform in the classical GSP is defined as the orthogonal base change w.r.t.\ an eigenbasis $\bU_G$ of a prescribed graph shift operator $\bS_G$ (e.g., the adjacency or Laplacian matrices). An interpretation is that each eigenvector of $\bS_G$ accounts for a level of signal smoothness quantified by its eigenvalue.  

Given a SAGS $\calA$, if we want to imitate the classical construction, we may copy the classical recipe and define the filter\footnote{As $\bU_G$ defines an orthogonal transformation that is norm preserving, we have $\mathfrak{c}_*(\mu) \in \calP(\mathbb{R}^n)$ for $\mu\in \calP(\mathbb{R}^n)$.}  $\phi: \calG_n \to M_n(\mathbb{R}), G \mapsto \bS_G \mapsto \bU_G$. Let $\mathfrak{c}=(\calA,\phi)$. Such a transform allows us to probe signal smoothness by incorporating probabilistic information. If $\calA$ is a constant SAGS, then the Fourier transform introduced in \cite{Ji22} is nothing but $e_{\mathfrak{c}}: \mathbb{R}^n \to \mathbb{R}^n$ (cf.\ \cref{thm:ftf}) when the notion of distributional graph signal is not yet introduced. 

From this explicit construction, we see the route to follow. Suppose a classical construction can be described by a function $f: \calG_n \to M_n(\mathbb{R})$. It also defines a filter if $c_*(\mu) \in \calP(\mathbb{R}^n)$ for $\mu \in \calP(\mathbb{R}^n)$, where $\mathfrak{c}=(\calA,f)$. For another important example, if $f: \calG_n \to M_n(\mathbb{R})$ is a filter such that $f(G)$ is a polynomial in (a prescribed GSO) $\bS_G$, then $\mathfrak{c}_*: \calP(\mathbb{R}^n) \to \calP(\mathbb{R}^n)$ is a \emph{convolution}. Similarly to Fourier transform, the notion of convolution introduced in \cite{Ji22} is nothing but $e_{\mathfrak{c}}$ for constant $\calA$.      
A special family of convolutions leads to the theory of sampling. Such a convolution takes the form $\rho: \calG_n \to M_n(\mathbb{R})$, where each $\rho(G)$ is the orthogonal projection matrix to the direct sum of a subcollection of eigenspaces of $\bS_{G}$. Let $\mathfrak{p} = (\calA,\rho)$. Inspired by \cite{Kaz67} and \cite{Ji22}, for $\epsilon>0$, a distributional graph signal $\mu$ is called \emph{$(\epsilon,\mathfrak{p})$-invariant} if $W(\mu, \mathfrak{p}_*(\mu))<\epsilon$. Recovery requires one to estimate such a $\mu$ based on its partial sampled observations. 

\begin{Example}
If $\calA$ is constant and $\mu = \delta_{\bx}, \bx \in \mathbb{R}^n$, then $\norm{\bx - e_{\mathfrak{p}}(\bx)} \leq W(\mu, \mathfrak{p}_*(\mu))$ by the Jensen inequality \cite{Dur19} and \cref{eg:imd}\ref{it:imd}. Therefore, if $\mu$ is $(\epsilon,\mathfrak{p})$-invariant, then $\bx$ is $(\epsilon,e_{\mathfrak{p}})$-bandlimited in the sense of \cite{Ji22}, where an explicit recovery scheme is given. If $\calA$ is locally constant, a brief discussion is given in Appendix~\ref{sec:rop}.
\end{Example}

Though it is impossible to discuss all important GSP concepts exhaustively, some essential ones have been covered. In the next section, we use numerical experiments to demonstrate how the framework of the paper can be applied in practice. 

\section{Experimental results} \label{sec:er}
\subsection{MNIST: examples of distributional graph signals} 
In this experiment, we showcase visualizations of distributional graph signals by summarising samples of each digit from $0$ to $9$ in the MNIST dataset\footnote{http://yann.lecun.com/exdb/mnist/} as a distributional signal. We preprocess the sample images by introducing i.i.d Gaussian noise to each pixel. The graph $G$ used is $28\times 28$ 2D-lattice.

We consider $2$ different approaches.
\begin{enumerate}[(I)]
    \item Edgewise Gaussian $\mu_E$ (abbreviated as ``Edgewise''): We learn from samples the joint Gaussian distribution of pairs of pixel values for each edge of the graph $G$. To draw a sample, we give $G$ an acyclic orientation with a single root. We draw a pixel value at the root using its marginal. For any directed edge, if the pixel value at the tail is already known, then the value at the head is drawn according to the conditional distribution derived from the joint distribution of the edge. The pixel values are averaged if a node is the head of multiple directed edges. The approach captures more refined pairwise signal relations in closed vicinity.   
    
    \item Joint Gaussian $\mu_G$ (abbreviated as ``Joint''): It is the joint Gaussian distribution of values at all the pixels that fits the samples. To draw a sample, we just draw from the joint distribution. The approach is based on a global perspective on the entire graph.
\end{enumerate}

\begin{figure}
    \centering
    \includegraphics[scale=0.55,trim=2cm 1.5cm 1.8cm 1cm,clip]{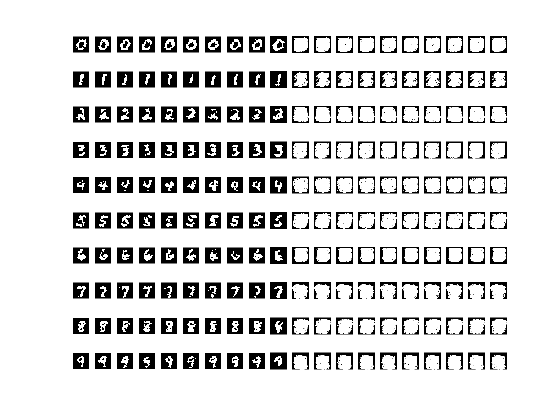}
    \caption{Samples drawn from $\mu_E$. The first half of the images are obtained by thresholding the second half of the images.}
    \label{fig:mnist2}
\end{figure}

\begin{figure}
    \centering
    \includegraphics[scale=0.55,trim=2cm 1.5cm 1.8cm 1cm,clip]{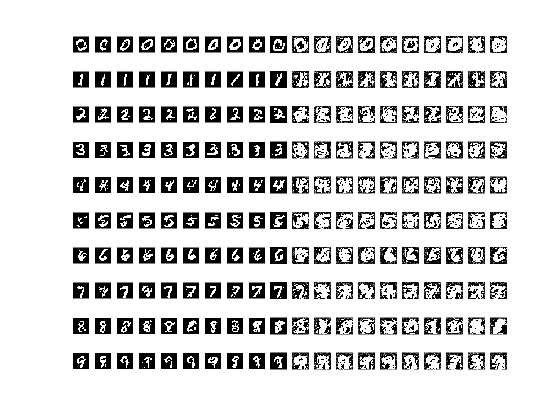}
    \caption{Samples drawn from $\mu_G$. The first half of the images are obtained by thresholding the second half of the images.}
    \label{fig:mnist3}
\end{figure}

We draw samples from both $\mu_E$ and $\mu_G$. From the sample images shown in (the right half of) \figref{fig:mnist2} and \figref{fig:mnist3}, we see that non of the approaches generate images with reasonable equality. For example, for $\mu_E$, the digits are not even recognizable. However, this does not necessarily mean that the distributions contain no useful information. We apply a thresholding function. The resulting samples are also shown in (the left half of) \figref{fig:mnist2} and \figref{fig:mnist3}. We see that now the digits are clearly recognizable. Moreover, $\mu_E$, the only distribution that leverages the graph structure, generates arguably the sharpest image of digits.

\vspace{.1in}

\begin{figure}
    \centering
    \includegraphics[scale=0.33]{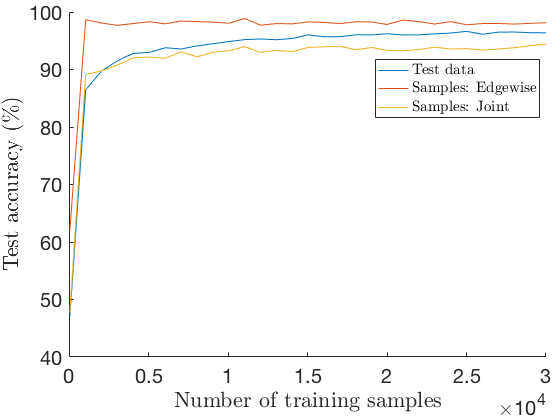}
    \caption{Test accuracy of the original test data and samples from the distributional graph signals.}
    \label{fig:sim}
\end{figure}

\begin{figure}
    \centering
    \includegraphics[scale=0.33]{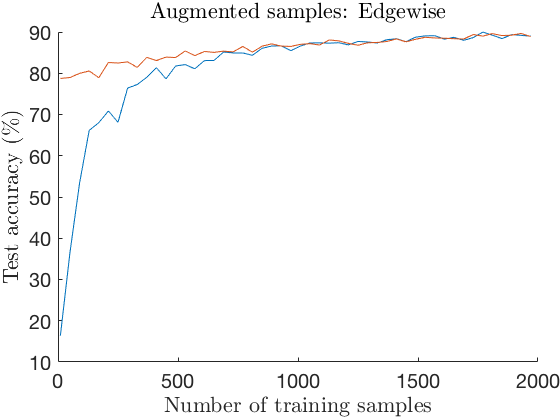}
    \includegraphics[scale=0.33]{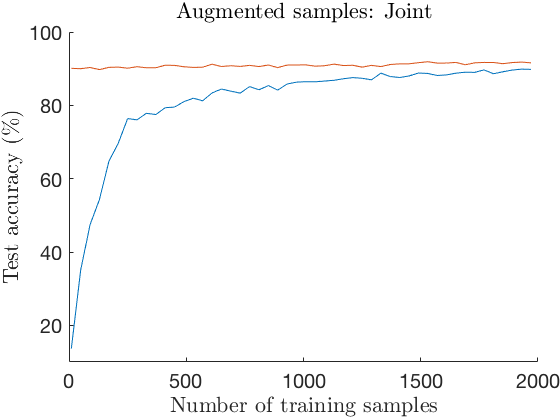}
    \caption{Test accuracy using training samples of small size (blue curve) and augmented training samples (red curve).}
    \label{fig:edg}
\end{figure}

We further investigate by resorting to the primary purpose that the dataset created: digit recognition. We take a base neural network model and perform the following two tasks. 
\begin{enumerate}[(a)]
    \item In the first task, we train the network with varying sizes of training sets. Then we test with the original test data (of size $10000$), as well as test data generated from distributional signals (Edgewise and Joint). The distributional signals are obtained using the original test data. The results are shown in \figref{fig:sim}. From the results, we see that the accuracy of samples from distributional signals: Edgewise is the highest in all the cases. This may suggest hidden statistical features might be captured by the distributional signals. 
    \item In the second task, we consider augmentation by distributional graph signals. We train the network with a small training set (of varying size $\leq 2000$). Moreover, we augment the dataset with $10000$ samples generated from distributional signals (Edgewise and Joint). Unlike the previous task, to get the distributional signals, we make use of a small portion of the original training dataset. We show the test results (in \figref{fig:edg}) on the original test dataset both with and without the augmentation. From the results, we notice that augmentation with distributional graph signals does significantly improve the test accuracy when the number of training samples is small. Moreover, using augmented samples: Joint has a better overall performance. 
    \end{enumerate}

\vspace{.1in}

\vspace{.1in}

The investigations suggest that samples from distributional signals: Edgewise might capture more details of the digits while using distributional signals: Joint can be more robust. To verify the last claim, we consider neural network adversarial attacks FGSM and PGD \cite{Goo15, Mad18}. More specifically, we use the original test dataset, while for the training dataset, we either use $10000$ samples from the original training dataset or $10000$ samples drawn from distributional signals: Joint. Test accuracies are shown in \cref{tab:poj}. We see that in general, the distributional approach can better resist adversarial attacks. 

\begin{table}[!htb]
\centering
\caption{} \label{tab:poj}
\begin{subtable}[t]{1\columnwidth}
\centering
\begin{tabular}{|c| c |c |c |} 
\hline
Perturbation & $0.1$ & $0.2$ & $0.3$  \\ 
\hline
Original & $34.3\%$ & $16.5\%$ & $11.9\%$  \\ 
\hline
Joint & $44.2\%$ & $25.2\%$ & $20.9\%$  \\ 
\hline
\end{tabular}
\caption{FGSM attack}
\label{fgsm}
\end{subtable}

\vspace{.1in}

\begin{subtable}[t]{1\columnwidth}
\centering
\begin{tabular}{|c| c |c| c| c| } 
\hline
Perturbation & $0.05$ & $0.1$ & $0.2$ & $0.3$  \\ 
\hline
Original & $72.8\%$ & $9.85\%$ & $0.16\%$ & $0.16\%$ \\ 
\hline
Joint & $96.0\%$ & $71.8\%$ & $4.74\%$ & $0\%$ \\ 
\hline
\end{tabular}
\caption{PGD attack}
\label{pgd}
\end{subtable}
\end{table}

\subsection{Weather dataset: filters and prediction} 
In this example, we consider filter learning for signal prediction. We use the US weather station network\footnote{http://www.ncdc.noaa.gov/data-access/} with 194 nodes, and they are connected by a $20$-NN graph $G$. Signals are temperature reading over a year. We want to learn a convolution filter $\bF$ in the normalized Laplacian $\widetilde{\bL_G}$ of degree up to $2$ that predicts temperature $4$ days or $7$ days in the future. In the setting of the paper, $\bF$ is $f(G)$ as in \cref{defn:gas}. We compare two approaches.

\begin{enumerate}[(a)]
    \item Classical GSP: we estimate $\bF$ that best predicts readings $4$ days (or $7$ days) in the future for $7$ consecutive days via a least mean square optimization.
    \item Distributional signals: we summarize readings in $7$ consecutive days as a (Guassian) distribution. The filter $\bF$ fits the distribution with the distribution of readings $4$ days (or $7$ days) in the future, by minimizing Wasserstein distance (\cref{eg:imd}). In particular, the variance of $7$ days reading at each station is taken into consideration.
\end{enumerate}

We perform the experiments for the readings in the $1$st half and $2$nd half of the years separately. We (uniformly) randomly sample a small fraction of groups of signals, with each group consisting of readings from $7$ consecutive days. For each group $\mathfrak{g}$ of readings, a filter $\bF_{\mathfrak{g}}$ is estimated using one of the two approaches described above. More specifically, let $a_{\mathfrak{g}}$ be the average reading (over all the stations) on the first day of the group $\mathfrak{g}$. An insight of \cref{eg:inn}\ref{it:agc} is that filters may change with signals. In this spirit, we may estimate $\bF_{\mathfrak{g}}$ that depends on $\mathfrak{g}$, e.g., the coefficients of $\bF_{\mathfrak{g}}$ (in $\widetilde{\bL_G}$) are themselves polynomials in $a_{\mathfrak{g}}$. In summary, given any number $a$, we can output a filter $\bF$ that is a degree $2$ polynomial in $\widetilde{\bL_G}$, whose coefficients are (learned and hence known) functions in $a$.

In testing, given any signal $\bx$, we compute $a_{\bx}$ as the average reading (over all the stations) of $\bx$. We hence obtain a filter $\widetilde{\bF_{\bx}}$ using $a_{\bx}$. The filter $\widetilde{\bF_{\bx}}$ is used for prediction and the performance is evaluated by the SNR of the predicted signal against the actual reading in the future. In the experiments below, we may consider either degree $2$ or degree $0$ polynomials in $a_{\mathfrak{g}}$ for filter coefficients. Degree $0$ is equivalent to the filter unchanged for different signals. 

\begin{figure}
    \centering
    \includegraphics[scale=0.3]{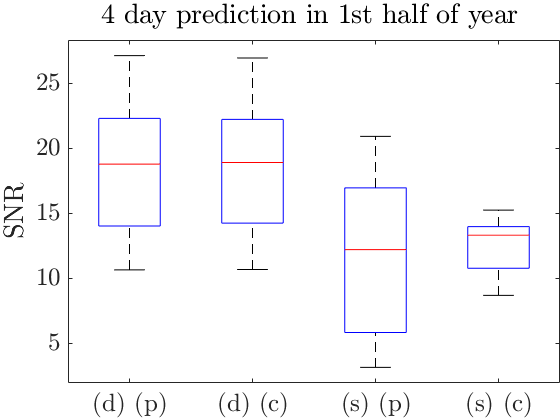}
    \includegraphics[scale=0.3]{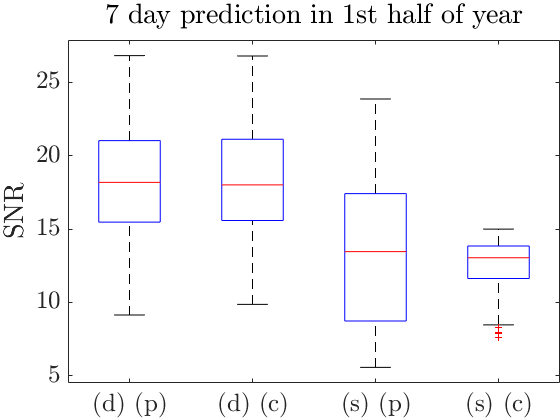}
    \includegraphics[scale=0.3]{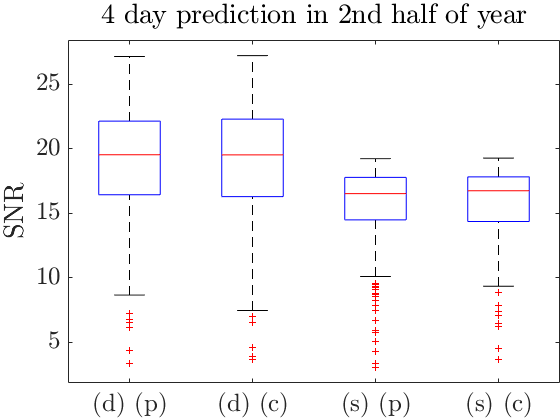}
    \includegraphics[scale=0.3]{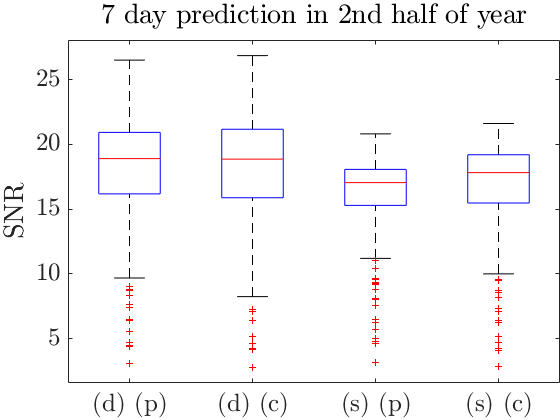}
    \caption{The performance of the four different approaches with $20\%$ of training samples.}
    \label{fig:snr1}
\end{figure}

In summary, we may propose approaches that consider distributional graph signals or classical (statistic) graph signals, denoted by (d) or (s) respectively for convenience. Moreover, the filter coefficients can either vary as polynomials in the mean of the signals or remain constant. The two situations are denoted by (p) and (c) for convenience. Altogether, we have four different combinations of approaches (d)(p), (d)(c), (s)(p), (s)(c). Their performance, with $20\%$ of training samples, is shown in \figref{fig:snr1}. We see that the distributional signal approaches have much better performance in all the cases. 

\begin{figure}
    \centering
    \includegraphics[scale=0.3]{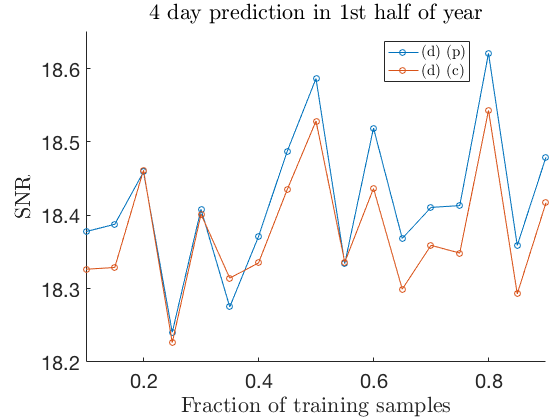}
    \includegraphics[scale=0.3]{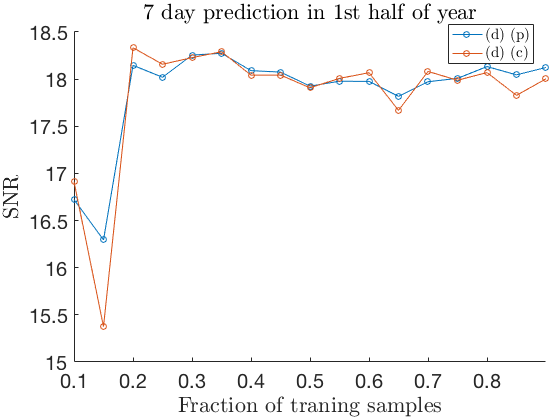}
    \includegraphics[scale=0.3]{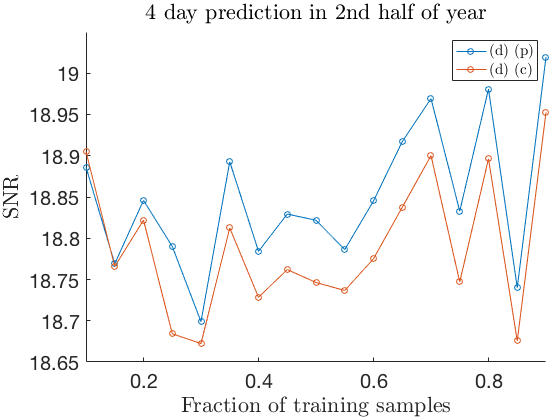}
    \includegraphics[scale=0.3]{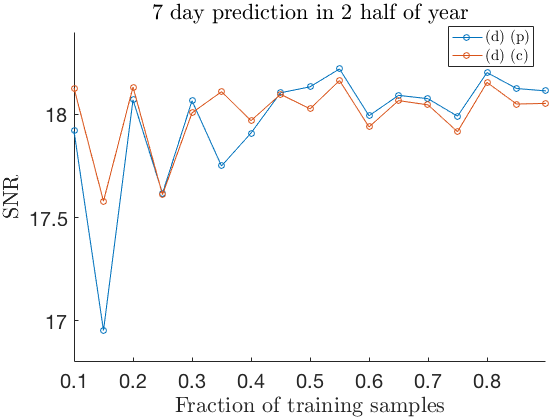}
    \caption{The average SNR comparison between (d)(p) and (d)(c) against different fractions of training samples.}
    \label{fig:snr2}
\end{figure}

To further compare (d)(p) and (d)(c), we vary the fraction of training samples and compute and record (in \figref{fig:snr2}) the average SNR for the two approaches. We see that for the $4$ days prediction when the predictions are supposed to be more accurate (as compared with $7$ days), (d)(p) is better than (d)(c) by a small margin but with a clear overall trend, i.e., it is preferable to let the filters change according to signals in the spirit of \cref{eg:inn}\ref{it:agc}. On the other hand, for the $7$ days prediction, (d)(p) and (d)(c) have comparable performance. In summary, using (d)(p) is at least as effective as the other approaches and can even be beneficial in some cases.  

\subsection{Brain ECoG dataset: anomaly detection}

In this experiment, we apply the framework of the paper to anomaly detection. We use the brain ECoG dataset.\footnote{\url{https://math.bu.edu/people/kolaczyk/datasets.html}} For each of the eight subjects in the dataset, there are $76$ sensors recording (normalized) brain ECoG signals in a time-series of $4000$ time-stamps. There are two signal types: pre-ict and ict signals. We consider ict signals abnormal. 

We segment the entire time-series into sub-intervals of size $10$ each. The $10$ time stamps can be modeled by the path graph $P$ on $10$ nodes. Suppose there is a connection $H$ among the sensors. Then there is the graph $G = H\times P$ of size $760$, with each node $v$ of $G=(V,E)$ corresponding to a pair $(s,t)$ where $s$ is a sensor and $t$ is a time-stamp. Different $H$ results in different $G$. A graph signal $\bx$ consists of sensor readings for $10$ consecutive time-stamps. 

We consider $\mathfrak{c}_j=(\calA_j,f_j), j=1,2$ with SAGS $\calA_j = (\mu_{j,\bx})_{\bx\in \mathbb{R}^{760}}$ defined as follows. We assume that for different subjects, their signals are disjoint, i.e., no two patients can have the same ECoG signal. Therefore, $\mathbb{R}^n$ is decomposed as $\mathbb{R}^n = \cup_{1\leq i\leq 8} C_i \cup C'$, where $C_i$ are all possible signals of the $i$-th subject and $C'$ is the complement of $\cup_{1\leq i\leq 8} C_i$ that plays no role in the problem. Therefore, we effectively consider locally constant SAGSs.

For any $G=H\times P$, let $\bL_G$ be its Laplacian. The filters $f_1, f_2$ are the high pass filter for the range from $730$ to $760$ w.r.t.\ $\bL_G$. Hence, $\mathfrak{c}_1$ and $\mathfrak{c}_2$ differ only in $\calA_1$ and $\calA_2$. To define $\calA_j$, first for $\bx \in C'$, let $\mu_{j,\bx}$ be supported on $G_0=H_0\times P$ for any fixed $H_0$ for convenience, as it is not used in the sequel. For $\bx\in C_i$, the empirical distribution of $H$ is estimated as in \cite{Ji22} Section VII D using $10\%$ of data as training samples. It is lengthy to give the details here, we just point out that in the estimation, one needs to specify a (graph) frequency range of $\bL_G$. We choose the frequency range from $0$ to $50$ for $\calA_1$ and from $50$ to $100$ for $\calA_2$.

We apply $\mathfrak{c}_{1,*}\boxplus\mathfrak{c}_{2,*}$ (\cref{sec:coe}) to each of the training set to obtain an empirical distribution in $\mathbb{R}^2$ and fit it with a mixed Gaussian with at most $3$ components. Though a mixed Gaussian may not be the best choice of distribution, we only need to know the positions of the peaks. For each subject, we randomly sample $10$ signals either all pre-ict or ict to form a signal test instance. The map $\mathfrak{c}_{1,*}\boxplus\mathfrak{c}_{2,*}$ is applied to the instance, and the resulting empirical distribution in $\mathbb{R}^2$ is again fitted with a mixed Gaussian distribution. It is compared with the base distributions. Examples for subject $1$ are shown in \figref{fig:prep1} and \figref{fig:ictp1}) (more are shown in the supplementary materials). We see that for both ict and pre-ict signals of each subject, the peak positions of Gaussian obtained from the test instances match well with those of the base distributions. 

\begin{figure}
    \centering
\includegraphics[width=1\columnwidth, height=.23\textheight]{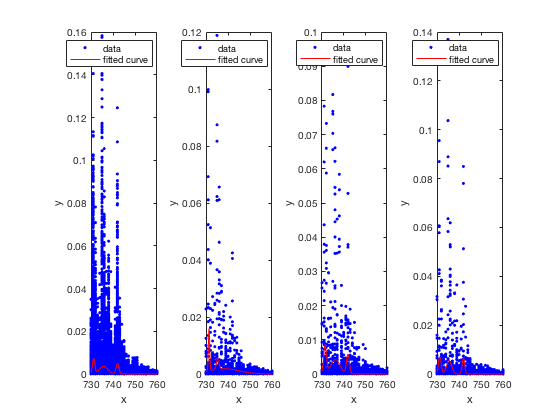}
    \caption{Pre-ict signals, subjects $1$: the base distribution and $3$ test instances.}
    \label{fig:prep1}
\end{figure}

\begin{figure}
    \centering
\includegraphics[width=1\columnwidth, height=.23\textheight]{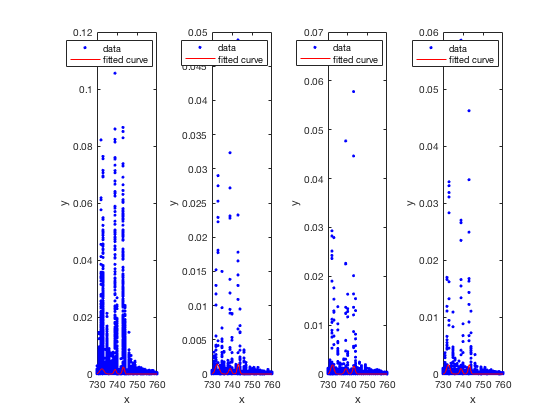}
    \caption{Ict signals, subjects $1$: the base distribution and $3$ test instances.}
    \label{fig:ictp1}
\end{figure}

\begin{figure}
    \centering
    \includegraphics[scale=0.35]{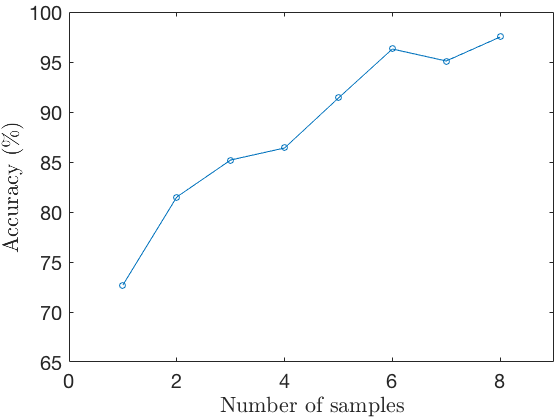}
    \caption{Accuracy of anomaly detection.}
    \label{fig:ano}
\end{figure}

The above observation suggests the following anomaly detection scheme. For the setup, we randomly choose a subject and a condition. Moreover, from the corresponding dataset (for the chosen patient and condition), we randomly draw a small number of samples ($\leq 8$). Let $\mu$ be the discrete distribution supported on the chosen samples. Using the method described earlier, we estimate the peak locations of the mixed Gaussian that fits $\mathfrak{c}_{1,*}\boxplus\mathfrak{c}_{2,*}(\mu)$ and compare to peaks locations of the $16$ base distributions. The comparison uses the Euclidean norm between the peak locations. The condition, either ict (abnormal) or pre-ict (normal), is declared using the corresponding condition of the base distribution models that are closest in average peak distance. We run the experiments for sample size $1,\ldots,8$ and compute the detection accuracy based on $200$ runs for each subject and condition. The results are shown in \figref{fig:ano}. We see that the accuracy increases rapidly if we increase the sample size. With $\geq 6$ samples, the accuracy is already $\approx 95\%$ or higher.   

\section{Conclusions} \label{sec:con}
In this work, we proposed a new approach to modeling and processing graph signals using distributional graph signals to account for signal stochasticity. Our proposed framework unifies existing approaches, provides a more flexible and realistic approach to modeling uncertain graph signals and graph topologies jointly, and has potential applications in various domains. The results of our experiments demonstrate the effectiveness of the proposed approach. We hope that this work can contribute to advancing the field of GSP by inspiring further research on the use of probability spaces in signal processing.

\appendices
\section{Proofs of theoretical results}\label{sec:pot}

\begin{proof}[Proof of \cref{thm:lkb}]
We first remark that as $K$ is compact, so is $\calP(K)$ by the Prokhorov theorem and the Skorokhod representation theorem \cite{Bil99}. 
Therefore, on $K$ and $\calP(K)$, any continuous function is also uniformly continuous. We first show that if $f_{\calA}$ is (uniformly) continuous, then the restriction of $\mathfrak{c}_*$ to $K$ is (uniformly) continuous. 

Consider $\bx_1,\bx_2 \in K$. Let $\gamma_{\bx_1,\bx_2}$ be a distribution on $M_n(\mathbb{R}) \times M_n(\mathbb{R})$ that realizes $W(f_{\calA}(\bx_1), f_{\calA}(\bx_2))$, i.e., 
\begin{align*}
    W(f_{\calA}(\bx_1), f_{\calA}(\bx_2))^2 = \int \norm{\bM_1-\bM_2}^2 \ud\gamma_{\bx_1,\bx_2}(\bM_1,\bM_2).
\end{align*}
The condition that $f_{\calA}$ is uniformly continuous means that: if $\bx_1,\bx_2$ are close enough in Euclidean distance, then the above integral can be arbitrarily small. Moreover, by uniform continuity, there is a uniform upper bound on the $W(f_{\calA}(\bx),0)$ for $\bx\in K$.

To estimate $W(\mathfrak{c}_*(\delta_{\bx_1}), \mathfrak{c}_*(\delta_{\bx_2}))$, consider 
\begin{align*}
M_n(\mathbb{R}) \times M_n(\mathbb{R}) \to \mathbb{R}^n\times \mathbb{R}^n, (\bM_1,\bM_2) \mapsto (\bM_1\bx_1, \bM_2\bx_2)
\end{align*}
and let $\gamma'_{\bx_1,\bx_2}$ be the pushforward probability distribution of $\gamma_{\bx_1,\bx_2}$ on $\mathbb{R}^n\times \mathbb{R}^n$. Moreover, based on the construction, the marginals of $\gamma'_{\bx_1,\bx_2}$ are $\mathfrak{c}_*(\delta_{\bx_1})$ and $\mathfrak{c}_*(\delta_{\bx_2})$. We have:
\begin{align*}
    & W(\mathfrak{c}_*(\delta_{\bx_1}), \mathfrak{c}_*(\delta_{\bx_2}))^2\\ \leq & \int \norm{\bz_1-\bz_2}^2 \ud\gamma'_{\bx_1,\bx_2}(\bz_1,\bz_2) \\
    = & \int \norm{\bM_1\bx_1 - \bM_2\bx_2}^2 \ud\gamma_{\bx_1,\bx_2}(\bM_1,\bM_2) \\
    \leq &\int  \norm{\bM_1\bx_1-\bM_2\bx_1}^2\ud\gamma_{\bx_1,\bx_2}(\bM_1,\bM_2) \\ & +\int  \norm{\bM_2\bx_1-\bM_2\bx_2}^2\ud\gamma_{\bx_1,\bx_2}(\bM_1,\bM_2) \\ \leq &(\int  \norm{\bM_1-\bM_2}^2\ud\gamma_{\bx_1,\bx_2}(\bM_1,\bM_2))(\norm{\bx_1}^2) \\ & + (\int \norm{\bM_2}^2df_{\calA}(\bx_2))( \norm{\bx_1-\bx_2}^2) \\ = &(\int  \norm{\bM_1-\bM_2}^2\ud\gamma_{\bx_1,\bx_2}(\bM_1,\bM_2))(\norm{\bx_1}^2) \\ & + W(f_{\calA}(\bx_2),0)^2\norm{\bx_1-\bx_2}^2.
\end{align*}
The last sum can be arbitrarily small if $\bx_1$ and $\bx_2$ are close enough because we have noticed that $W(f_{\calA}(\bx),0)$ is uniformly bounded. Moreover, it is independent of the location of $\bx_1$ in $K$. Therefore, $\mathfrak{c}_*$ is uniformly continuous when restricted to $K$. This further implies that for $\epsilon > 0$, there is $B_{\epsilon}$ depending only on $\epsilon$ such that if $\bx_1,\bx_2 \in K$ satisfy $\norm{\bx_1-\bx_2} \geq \epsilon$, then $W(\mathfrak{c}_*(\delta_{\bx_1}), \mathfrak{c}_*(\delta_{\bx_2}))^2 \leq B_{\epsilon}\norm{\bx_1-\bx_2}^2$. Moreover, there is a $C_{\epsilon}$ also depending only on $\epsilon$ such that if $\norm{\bx_1-\bx_2} < \epsilon$, then $W(\mathfrak{c}_*(\delta_{\bx_1}), \mathfrak{c}_*(\delta_{\bx_2}))^2\leq C_{\epsilon}$. As $\epsilon \to 0$, $C_{\epsilon} \to 0$. We also remark that based on the expression, if the compact set $K$ is contained in the ball of radius $R$ (centered at the origin) in $\mathbb{R}^n$, then  $B_{\epsilon} = O(R)$. This will be used in the next proof.  

Consider general $\mu,\mu'$ on $K \subset \mathbb{R}^n$. Let $\eta$ be a distribution on $\mathbb{R}^n\times \mathbb{R}^n$ that realizes $W(\mu,\mu')$ and $\gamma_{\bx_1,\bx_2}$ be defined earlier for $(\bx_1,\bx_2) \in \mathbb{R}^n\times \mathbb{R}^n$. Define a distribution $\eta'$ on $\mathbb{R}^n\times \mathbb{R}^n$ by the following integral equation. For any compactly supported continuous function $g$ on $\mathbb{R}^n\times \mathbb{R}^n$, $\eta'$ satisfies:
\begin{align*}
    & \int g(\bw_1,\bw_2) \ud\eta'(\bw_1,\bw_2) \\ = &\int \int g(\bw_1,\bw_2)\ud\gamma_{\bx_1,\bx_2}(\bw_1,\bw_2)\ud\eta(\bx_1,\bx_2).
\end{align*}

We verify that the marginals of $\eta'$ are $\mathfrak{c}_*(\mu)$ and $\mathfrak{c}_*(\mu')$ respectively. Let $p: \mathbb{R}^n\times \mathbb{R}^n \to \mathbb{R}^n$ be the projection to either component. Consider any compactly supported continuous function $g$ on $\mathbb{R}^n$. We have
\begin{align*}
    & \int g(\bw_1)\ud p_*(\eta')(\bw_1)  =  \int g(\bw_1) \ud\eta'(\bw_1,\bw_2) \\
     = &\int \int g(\bw_1)d\gamma_{\bx_1,\bx_2}(\bw_1,\bw_2)\ud\eta(\bx_1,\bx_2) \\ 
     = &\int \int g(\bw_1)d\mathfrak{c}_*({\delta_{\bx_1}})(\bw_1)\ud\eta(\bx_1,\bx_2) \\ = & \int \int g(\bw_1)d\mathfrak{c}_*({\delta_{\bx_1}})(\bw_1)\ud\mu(\bx_1)\\  = &\int g(\bw_1) \ud\mathfrak{c}_*(\mu)(\bw_1). 
\end{align*}
This proves the claim.

For any $\epsilon > 0$, we estimate 
\begin{align*}
    & W(\mathfrak{c}_*(\mu),\mathfrak{c}_*(\mu'))^2  \leq \int \norm{\bw_1-\bw_2}^2 \ud\eta'(\bw_1,\bw_2)  \\
    = & \int \int \norm{\bw_1-\bw_2}^2\ud\gamma_{\bx_1,\bx_2}(\bw_1,\bw_2)\ud\eta(\bx_1,\bx_2) \\ = & \int_{\norm{\bx_1-\bx_2}\geq \epsilon} \int \norm{\bw_1-\bw_2}^2\ud\gamma_{\bx_1,\bx_2}(\bw_1,\bw_2)\ud\eta(\bx_1,\bx_2) \\ + & \int_{\norm{\bx_1-\bx_2} < \epsilon} \int \norm{\bw_1-\bw_2}^2\ud\gamma_{\bx_1,\bx_2}(\bw_1,\bw_2)\ud\eta(\bx_1,\bx_2) \\
    \leq & \int B_{\epsilon}\norm{\bx_1-\bx_2}^2 \ud\eta(\bx_1,\bx_2) +  \int C_{\epsilon} \ud\eta(\bx_1,\bx_2) \\ = & B_{\epsilon}W(\mu,\mu') + C_{\epsilon}.
\end{align*}
Therefore, as long as $\epsilon$ (chosen first) is small enough and $W(\mu,\mu')$ is small enough, $W(\mathfrak{c}_*(\mu),\mathfrak{c}_*(\mu'))^2$ can be arbitrarily small. This proves the theorem.  
\end{proof}

\begin{proof}[Proof of \cref{thm:imi}]
The structure of the proof follows that of the proof of \cref{thm:lkb}. Following the argument of \cref{thm:lkb} and using Lipschitz continuity of $f_{\calA}$, for any compact subset $K$ of $\mathbb{R}^n$, there is a $B_K$ depending only on $K$ such that $W(\mathfrak{c}_*(\delta_{\bx_1}), \mathfrak{c}_*(\delta_{\bx_2}) )^2\leq B_K\norm{\bx_1-\bx_2}^2$ for any $\bx_1,\bx_2\in K$. Moreover, if $K$ is contained in the ball centered at the origin with radius $R$, then $B_K = O(R)$. 

Let $\mu \in \mathcal{P}(\mathbb{R}^n)$ have finite $6$-th moment and $\epsilon>0$. By the Markov inequality, there is a closed ball $K_{\epsilon}$ (centered at the origin) with radius $R_{\epsilon} = o(1/\epsilon)$ such that $\int_{\bx\notin K_{\epsilon}}\norm{\bx}^2 d\mu(\bx) \leq \epsilon$. Let $K'_{\epsilon}$ be the ball (centered at the origin) with radius $2R_{\epsilon}$. Consider any $\mu' \in \mathcal{P}(\mathbb{R}^n)$ such that $W(\mu,\mu') \leq \epsilon$. Let $\eta$ be the distribution on $\mathbb{R}^n\times \mathbb{R}^n$ that realizes $W(\mu,\mu')$. We estimate: 
\begin{align*}
    \epsilon \geq & \int \norm{\bx_1-\bx_2}^2\ud\eta(\bx_1,\bx_2) \\
    \geq & \int_{\bx_1 \in K_{\epsilon}, \bx_2\notin K'_{\epsilon}}\norm{\bx_1-\bx_2}^2\ud\eta(\bx_1,\bx_2) \\ & + \int_{\bx_1 \notin K_{\epsilon}, \bx_2\notin K'_{\epsilon}}\norm{\bx_1-\bx_2}^2\ud\eta(\bx_1,\bx_2)\\
    \geq & \frac{1}{4}\int_{\bx_1 \in K_{\epsilon}, \bx_2\notin K'_{\epsilon}}\norm{\bx_2}^2\ud\eta(\bx_1,\bx_2) \\ & + \frac{1}{2}\int_{\bx_1\notin K_{\epsilon}, \bx_2\notin K'_{\epsilon}}\norm{\bx_2}^2\ud\eta(\bx_1,\bx_2)\\ 
    & - \int_{\bx_1\notin K_{\epsilon}, \bx_2\notin K'_{\epsilon}}\norm{\bx_1}^2\ud\eta(\bx_1,\bx_2) \\
    \geq & \frac{1}{4}\int_{\bx_2\notin K'_{\epsilon}}\norm{\bx_2}^2\ud\eta(\bx_1,\bx_2) - \int_{\bx_1\notin K_{\epsilon}}\norm{\bx_1}^2\ud\eta(\bx_1,\bx_2) \\
    = & \frac{1}{4} \int_{\bx_2 \notin K'_{\epsilon}}\norm{\bx_2}^2d\mu'(\bx_2) - \int_{\bx_1\notin K_{\epsilon}}\norm{\bx_1}^2d\mu(\bx_1).
\end{align*}
Therefore, $\int_{\bx\notin K'_{\epsilon}}\norm{\bx}^2d\mu'(\bx)\leq 8\epsilon$.

Since it is assumed that $f_{\calA}$ is Lipschitz, there is $B_0$ such that $W(f_{\calA}(\bx_1),f_{\calA}(\bx_1))^2 \leq B_0\norm{\bx_1-\bx_2}^2$ for every $\bx_1,\bx_2\in \mathbb{R}^n$. For $\mu'\in \mathcal{P}(\mathbb{R})^n$ such that $W(\mu,\mu')\leq \epsilon$, choose $K'_{\epsilon}$ and hence $B_{K'_{\epsilon}} = o(1/\epsilon)$ as earlier in the proof. Moreover, let $\eta$ and $\gamma_{\bx_1,\bx_2}, \bx_1,\bx_2\in \mathbb{R}^n$ be as in the proof of \cref{thm:lkb}, the same estimation yields:
\begin{align*}
    & W(\mathfrak{c}_*(\mu),\mathfrak{c}_*(\mu'))^2 \\
    \leq & \int \int \norm{\bw_1-\bw_2}^2\ud\gamma_{\bx_1,\bx_2}(\bw_1,\bw_2)\ud\eta(\bx_1,\bx_2) \\
    = & \int_{\bx_1\in K'_{\epsilon}, \bx_2 \in K'_{\epsilon}} \int \norm{\bw_1-\bw_2}^2\ud\gamma_{\bx_1,\bx_2}(\bw_1,\bw_2)\ud\eta(\bx_1,\bx_2) \\ + & \int_{\bx_1\in K'_{\epsilon}, \bx_2 \notin K'_{\epsilon}} \int \norm{\bw_1-\bw_2}^2\ud\gamma_{\bx_1,\bx_2}(\bw_1,\bw_2)\ud\eta(\bx_1,\bx_2) \\  + & \int_{\bx_1\notin K'_{\epsilon}, \bx_2 \in K'_{\epsilon}} \int \norm{\bw_1-\bw_2}^2\ud\gamma_{\bx_1,\bx_2}(\bw_1,\bw_2)\ud\eta(\bx_1,\bx_2) \\
    = & \int_{\bx_1\in K'_{\epsilon}, \bx_2 \in K'_{\epsilon}} \int \norm{\bw_1-\bw_2}^2\ud\gamma_{\bx_1,\bx_2}(\bw_1,\bw_2)\ud\eta(\bx_1,\bx_2) \\ & + \int_{\bx_1\in K'_{\epsilon}, \bx_2 \notin K'_{\epsilon}} W(f_{\calA}(\bx_1),f_{\calA}(\bx_1))^2\ud\eta(\bx_1,\bx_2) \\ & + \int_{\bx_1\notin K'_{\epsilon}, \bx_2 \in K'_{\epsilon}} W(f_{\calA}(\bx_1),f_{\calA}(\bx_1))^2\ud\eta(\bx_1,\bx_2) \\ \leq & B_{K'_{\epsilon}}\int \norm{\bx_1-\bx_2}^2\ud\eta(\bx_1,\bx_2) \\ & + \int_{\bx_2 \notin K'_{\epsilon}} 4B_0\norm{\bx_2}^2\ud\eta(\bx_1,\bx_2) \\
    & + \int_{\bx_1 \notin K'_{\epsilon}} 4B_0\norm{\bx_1}^2\ud\eta(\bx_1,\bx_2) \\
    \leq & B_{K'_{\epsilon}}W(\mu,\mu') \\
    & + 4B_0\Big( \int_{\bx_2\notin K'_{\epsilon}}\norm{\bx_2}^2\ud\mu'(\bx_2) + \int_{\bx_1\notin K'_{\epsilon}}\norm{\bx_1}^2\ud\mu(\bx_1) \Big)  \\
    \leq & (B_{K'_{\epsilon}}+32B_0)\epsilon.
\end{align*}
As $(B_{K'_{\epsilon}}+32B_0)\epsilon = o(1)$, the distance $W(\mathfrak{c}_*(\mu),\mathfrak{c}_*(\mu'))$ can be arbitrarily small as long as $\epsilon \to 0$ and $W(\mu,\mu') < \epsilon$. The theorem is proved. 
\end{proof}

\begin{proof}[Proof of \cref{thm:ftf}]
For $\mathfrak{c} =(\calA,f)$, we first show that $e_{\mathfrak{c}}$ is measurable. Let $C$ be any compact subset of $\mathbb{R}^n$ and $\mu_C$ be the uniform distribution on $C$. It induces the measure $\calA^*(\mu_C)$ on $Y = \mathbb{R}^n\times M_n(\mathbb{R})$. Moreover, it is easy to verify that $p_*\circ \calA^*(\mu_C) = \mu_C$, where $p: Y \to \mathbb{R}^n$ is the projection.

We view $Y$ as the sample space with probability distribution $\calA^*(\mu_C)$ and measurable functions $p, f$ as random variables. Let $e_C: \mathbb{R}^n \to \mathbb{R}^n$ be the associated conditional expectation. By the construction, we have $e_C = e_{\mathfrak{c}}$ on $C$ and $e_C = 0$ on the complement $\mathbb{R}^n\backslash C$. Moreover, $e_C$ is measurable w.r.t.\ the measure $\mu_C$. However, as $\mu_C$ is uniform, $e_C$ is also measurable w.r.t.\ the Lebesgue measure.  

Let $C_1 \subset C_2 \subset \ldots \subset C_i \subset \ldots$ be a sequence of compact subsets of $\mathbb{R}^n$ such that $\cup_{i\geq 1}C_i = \mathbb{R}^n$. Then $(e_{C_i})_{i\geq 1}$ is a sequence of measurable functions whose pointwise limit is $e_{\mathfrak{c}}$. Therefore, $e_{\mathfrak{c}}$ is also measurable. 

As a consequence, given any distribution $\mu$ on $\mathcal{P}(\mathbb{R}^n)$, pushforward of $e_{\mathfrak{c}}$ induces a distribution $e_{\mathfrak{c},*}(\mu)$. We need to show that $e_{\mathfrak{c},*}(\mu) \in \calP(\mathbb{R}^n)$ in order to claim that $e_{\mathfrak{c},*}$ is well defined as a map $\mathcal{P}(\mathbb{R}^n) \to \mathcal{P}(\mathbb{R}^n)$. Consider the Jensen inequality \cite{Dur19}:
\begin{align} \label{eq:nmx}
    \norm{\mathbb{E}_{G\sim \mu_{\bx}}f(G)(\bx)}^2\leq \mathbb{E}_{G\sim \mu_{\bx}}\norm{f(G)(\bx)}^2, \bx\in \mathbb{R}^n.
\end{align}
For any $\mu \in \mathcal{P}(\mathbb{R}^n)$, to show that $e_{\mathfrak{c},*}(\mu) \in \mathcal{P}(\mathbb{R}^n)$, it suffices to check that 
\begin{align*}
    \int \norm{\mathbb{E}_{G\sim \mu_{\bx}}f(G)(\bx)}^2 d\mu(\bx) < \infty,
\end{align*}
as finiteness of mean follows from that of $\mathfrak{c}_*(\mu)$ and linearity of expectation.
However, by (\ref{eq:nmx}), the left-hand side is bounded by 
\begin{align*}
    \int \mathbb{E}_{G\sim \mu_{\bx}}\norm{f(G)(\bx)}^2 d\mu(\bx) < \infty,
\end{align*}
due to the assumption that $\mathfrak{c}_*(\mu) \in \mathcal{P}(\mathbb{R}^n)$ and its $2$nd moment is finite. 

To show the claimed inequality, let $S$ be a subset $\mathbb{R}^n$ and $\mu \in \calP(\mathbb{R}^n)$ be supported on $S$. Consider $\eta$ the pushforward measure of $\calA^*(\mu)$ on $\mathbb{R}^n\times \mathbb{R}^n$ via the map: $Y\to \mathbb{R}^n\times \mathbb{R}^n, (\bx,G)\mapsto (e_{\mathfrak{c}}\circ p(\bx,G),f(G)\bx) = (e_{\mathfrak{c}}(\bx),f(G)\bx)$. The marginals of $\eta$ are 
\begin{align*}
(e_{\mathfrak{c}}\circ p)_*(\calA^*(\mu)) = e_{\mathfrak{c},*}(\mu) \text{ and } f_*\circ \calA^*(\mu) = \mathfrak{c}(\mu)
\end{align*}
respectively. Therefore,   
\begin{align*}
    & W(e_{\mathfrak{c},*}(\mu), \mathfrak{c}(\mu))^2 \\
    \leq & \int \norm{\bx_1-\bx_2}^2 \ud\eta(\bx_1,\bx_2) \\
    = & \int \norm{e_{\mathfrak{c}}(\bx)-f(G)(\bx)}^2 \ud \calA^*(\mu)(\bx,G) \\
    = & \mathbb{E}_{(\bx,G) \sim \calA^*(\mu)}\norm{e_{\mathfrak{c}}(\bx) -f(G)(\bx)}^2 \\
    \leq & \mathbb{E}_{(\bx,G) \sim \calA^*(\mu)}\norm{g(\bx) -f(G)(\bx)}^2,
\end{align*}
The last inequality holds as $e_{\mathfrak{c}}$ is the conditional expectation (up to a set of measure $0$) w.r.t.\ $\calA^*(\mu)$ on $Y$ \cite{Kol56}. 

To estimate the right-hand-side, we have 
\begin{align*}
&\mathbb{E}_{(\bx,G) \sim \calA^*(\mu)}\norm{g(\bx) -f(G)(\bx)}^2 \\ \leq &\sup_{\bx\in S} \mathbb{E}_{(\bx,G) \sim \calA^*(\delta_{\bx})}\norm{g(\bx) -f(G)(\bx)}^2.
\end{align*}
As $g(\bx)$ is independent of $G$, we have 
\begin{align*}
& \mathbb{E}_{(\bx,G) \sim \calA^*(\delta_{\bx})}\norm{g(\bx) -f(G)(\bx)}^2 \\ = & W\big(\delta_{g(\bx)},\mathfrak{c}(\delta_{\bx})\big)^2 = W\big(g_*(\delta_{\bx}),\mathfrak{c}(\delta_{\bx})\big)^2\\ \leq &\sup_{\text{supp}(\nu) \subset S} W\big(g_*(\nu), \mathfrak{c}(\nu)\big)^2.
\end{align*}
The result follows.
\end{proof}

\begin{proof}[Proof of \cref{lem:cas}]
Suppose $\gamma$ is a distribution on $M_n(\mathbb{R})\times M_n(\mathbb{R})$ that realizes $W(f_{i,\calA_i}(\bx),f_{\calA}(\bx))$. We estimate
\begin{align*}
    & W(f_{i,\calA_i}(\bx),f_{\calA}(\bx))^2\norm{\bx}^2 \\
    = & \int \norm{\bM_1-\bM_2}^2\norm{\bx}^2 \ud \gamma(\bM_1,\bM_2) \\
    \geq & \big(\int \norm{(\bM_1-\bM_2)\bx} \ud \gamma(\bM_1,\bM_2) \big)^2 \\
    \geq & \norm{\int (\bM_1-\bM_2)\bx\ud\gamma(\bM_1,\bM_2)}^2 \\
    = & \norm{\int \bM_1\bx\ud\gamma(\bM_1,\bM_2)-\int \bM_2\bx\ud\gamma(\bM_1,\bM_2)}^2 \\
    = & \norm{\int \bM_1\bx\ud f_{i,\calA_i}(\bx)(\bM_1)-\int \bM_2\bx\ud f_{\calA}(\bx)(\bM_2)}^2 \\
    = & \norm{e_{\mathfrak{c}_i}(\bx)-e_{\mathfrak{c}}(\bx)}^2. 
\end{align*}
Therefore, if $f_{i,\calA_i}(\bx) \to f_{\calA}(\bx), i\to \infty$, then $e_{\mathfrak{c}_i}(\bx) \to e_{\mathfrak{c}}(\bx)$. 
\end{proof}

\begin{proof}[Proof of \cref{lem:tam}]
To show $\mathfrak{c}_{1,*}\boxplus\mathfrak{c}_{2,*}$ is well-defined, we want to prove that $\mathfrak{c}_{1,*}\boxplus\mathfrak{c}_{2,*}(\mu) \in \calP(\mathbb{R}^n)$ if $\mu \in \calP(\mathbb{R}^n)$. Let $g$ be the function $\bx \mapsto \norm{\bx}^2$, and we have
\begin{align*}
&\mathfrak{c}_{1,*}\boxplus\mathfrak{c}_{2,*}(\mu)(g) = \\ &\int \int \norm{f_1(G_1)(\bx)+f_2(G_2)(\bx)}^2 \ud\nu_{1,\bx}\times \nu_{2,\bx}(G_1,G_2)\ud \mu(\bx) \\
& \leq \int \int 2\norm{f_1(G_1)(\bx)}^2 \ud\nu_{1,\bx}\times \nu_{2,\bx}(G_1,G_2)\ud \mu(\bx) \\
& + \int \int 2\norm{f_2(G_2)(\bx)}^2 \ud\nu_{1,\bx}\times \nu_{2,\bx}(G_1,G_2)\ud \mu(\bx) \\
& = 2\mathfrak{c}_{1,*}(\mu)(g) + 2\mathfrak{c}_{2,*}(\mu)(g).  
\end{align*}
Therefore, $\mathfrak{c}_{1,*}\boxplus\mathfrak{c}_{2,*}(\mu)$ has finite variance as both $f_1$ and $f_2$ are finites. Similarly, if we choose $g$ to be the identity function, we see that $\mathfrak{c}_{1,*}\boxplus\mathfrak{c}_{2,*}(\mu)$ has finite mean. Hence, $\mathfrak{c}_{1,*}\boxplus\mathfrak{c}_{2,*}(\mu)$ is in $\calP(\mathbb{R}^n)$. 

To verify the algebraic identity, we compute
\begin{align*}
     & e_{r\mathfrak{c}_1+\mathfrak{c}_2}(\bx)
    = \int \by \ud r\mathfrak{c}_{1,*}\boxplus\mathfrak{c}_{2,*}(\delta_{\bx})(\by) \\
    = &\int rf_1(G_1)(\bx)+f_2(G_2)(\bx) d\nu_{1,\bx}\times \nu_{2,\bx}(G_1,G_2) \\ 
    = &\int rf_1(G_1)(\bx)d\nu_{1,\bx}(G_1)\\
    &+ \int f_2(G_2)(\bx)d\nu_{2,\bx}(G_2)\\
    = & e_{r\mathfrak{c}_1}(\bx)+e_{\mathfrak{c}_2}(\bx).
\end{align*}
\end{proof}

\section{A category theoretical perspective}\label{sec:act}

Category theory \cite{Hun03} is a branch of mathematics that deals with the abstract study of structures and relationships between objects. It provides a framework for organizing mathematical concepts and objects. Category theory aims to identify common patterns and structures across different mathematical disciplines and provide a unified language for talking about these structures. 

A category $\calC$ consists of the following entities:
\begin{itemize}
    \item A class $ob(\calC)$, the \emph{objects} of $\calC$.
    \item For every pair objects $C_1,C_2$, a class $mor(C_1,C_2)$ of \emph{morphisms} from $C_1$ to $C_2$. 
    \item For any triple of objects $C_1,C_2,C_3$, there is the composition $\circ: mor(C_1,C_2)\times mor(C_2,C_3) \to mor(C_1,C_3)$ expressed as $\circ(f,g) = g\circ f$ such that 
    \begin{enumerate}[(a)]
        \item $f\circ (g \circ h) = (f\circ g)\circ h$. 
        \item For each object $C$, there is the identity morphism $1_C \in mor(C,C)$ such that $f\circ 1_C = f = 1_C\circ f$. 
    \end{enumerate}
\end{itemize}
The most relevant category to traditional GSP is $\text{Vect}_{\mathbb{R}}$, the category of finite dimensional vector spaces. In $\text{Vect}_{\mathbb{R}}$, the objects are finite dimensional $\mathbb{R}$-vector spaces, and the morphisms between a pair of vector spaces are the linear transformations between them. More generally, we have Meas the category of measurable spaces. The morphisms between two measurable spaces are measurable functions.

The framework of the paper can also be described using a category $\calC$. We highlight some essential ideas. The objects are measurable spaces. A morphism (up to certain equivalence) $\mathfrak{c}=(Y,f_1,f_2)$ between two objects $X_1,X_2$ consists of a measurable space $Y$ and measurable functions $f_1: Y \to X_1$ and $f_2: Y \to X_2$ such that the following holds (cf.\ \cite{Ji22} Section V): 
\begin{itemize}
    \item For each $x\in X_1$, there a probability distribution $\mu_x$ on $f_1^{-1}(x)$. The collection of \emph{fiberwise} distributions $(\mu_x)_{x\in X_1}$ induces for any probability measure $\mu$ on $X_1$, a probability measure on $f_1^*(\mu)$ on $Y$.
    \item Let ${f_2}_*$ be the pushforward map of probability measures on $Y$. The composition ${f_2}_*\circ f_1^*$ is well-defined as a map $\calP(X_1) \to \calP(X_2)$.
\end{itemize}
In the setup, the primary example of $X_1$ is the graph signal space $\mathbb{R}^n$. Graph structural information is encoded in $f_1: Y \to \mathbb{R}^n$ and $(\mu_x)_{x\in \mathbb{R}^n}$, when $Y$ consists of pairs $(x,G)$ with $G$ a graph of size $n$. The notion of fiberwise distributions $(\mu_x)_{x\in \mathbb{R}^n}$ corresponds to that of SAGS in \cref{sec:bay}. On the other hand, $f_2$ is related to signal transformation including filtering. 

As category theory is out of the scope of the paper, details on the categorical perspective can be found in \cite{Ji23c}.

\section{Remarks on piecewise linear functions} \label{sec:rop}

As we have seen in \cref{eg:ici} that for $\mathfrak{p}=(\calA,f)$, if $\calA$ is locally constant, then $e_{\mathfrak{p}}$ is piecewise linear a.e. Let $e_{\mathfrak{p}}$ consist of linear transformations $(\bP_i)_{i\geq 1}$. To simplify the discussion, we assume that each $\bP_i$ is a projection to an $m$-dimensional subspace $\calW_i$ of $\mathbb{R}^n$ for $m<n$. We want to discuss sampling and recovery with this setup. A rigorous discussion requires the theory of Grassmannians to parametrize linear spaces\cite{Mil74}, which is out of the scope. We content to explain the main idea.  

In classical GSP when there is only a single linear projection $\bP$ to ($m$ dimensional) $\calW$, then sampling and recovery amount to find a set of $m$ coordinates corresponding to $V'\subset V$, and identify the intersection of $\calW$ with the signal space $\calS$ with fixed observation on $V'$. 

This approach does not work for a set of projections $(\bP_i)_{i\geq 1}$ as above. Assume that there is an index $j$ such that $\calW_j\cap \calS = \{\bx\}$, which we want to identify. The challenge is that $\calS\cap \calW_i$ is usually non-empty for any $i\geq 1$. Therefore, it is not possible to find $\bx$ based on the partial observations at $V'$.  

However, the issue can be resolved by enlarging $V'$ by including one more sample. For the new $V'$ and $\calS$, it is usually true that $\calS \cap \calW_i = \emptyset$ by dimension counting, except for the single index $j$ that is known (a priori) to satisfy $\calW_j \cap \calS = \{\bx\}$. 

Though we have been vague in the claims by using ``usually'' a few times, it is possible to make them precise by stating ``non-empty open set in the Grassmannian manifold''. However, our key message here is that in almost any case, it is necessary and sufficient to find $m+1$ samples. 

\bibliographystyle{IEEEtran}
\bibliography{IEEEabrv,StringDefinitions,allref}

\end{document}